\documentclass[aps,pre,twocolumn,groupedaddress,showpacs,floatfix]{revtex4}
\usepackage{graphicx}
\usepackage{amsmath}
\usepackage{graphicx}
\usepackage{graphics}
\usepackage{amsfonts}
\usepackage{amssymb}

\usepackage[utf8]{inputenc}
\begin{document}

\title{Numerical study of a model for non-equilibrium wetting}

\author{A. C. Barato,$^1$ H. Hinrichsen,$^1$  and M. J. de Oliveira$^2$}

\affiliation{$^1$Fakult{\"a}t f{\"u}r Physik und Astronomie, Universit{\"a}t W{\"u}rzburg, Am Hubland, 97074 W{\"u}rzburg, Germany\\
	$^2$Instituto de F\'\i sica, Universidade de S\~ao Paulo, Caixa Postal 66318, 05315-970 S\~ao Paulo, S\~ao Paulo, Brazil}

\begin{abstract}
We revisit the scaling properties of a model for non-equilibrium wetting [Phys. Rev. Lett. {\bf 79}, 2710 (1997)], correcting previous estimates of the critical exponents and providing a complete scaling scheme. Moreover, we investigate a special point in the phase diagram, where the model exhibits a roughening transition related to directed percolation. We argue that in the vicinity of this point evaporation from the middle of plateaus can be interpreted as an external field in the language of directed percolation. This analogy allows us to compute the crossover exponent and to predict the form of the phase transition line close to its terminal point.
\end{abstract}


\maketitle
\parskip 1mm 

\section{Introduction}

While wetting of surfaces at or near thermal equilibrium is well understood~\cite{diet86},
the study of wetting phenomena \textit{far} from equilibrium is a challenging new
field. In the past decade there have been numerous theoretical studies addressing 
the question whether non-equilibrium conditions may lead to different physical
phenomena near the wetting transition. Most of these studies are based on particular 
lattice models~\cite{hinr97, MW, kiss05} or phenomenological Langevin equations~\cite{tu97}. 
They all have in common that the wetting layer is modelled by a non-equilibrium growth process of a
$d$-dimensional interface combined with a hard-core wall which represents the
surface of the substrate.

The theoretical interest in non-equilibrium wetting stems from the fact
that various scale-invariant properties are found to be universal, i.e., they are dictated
by the symmetries of the model irrespective of microscopic details. For a
growth process without a substrate (free interface), the most prominent universality classes are
the Edwards Wilkinson (EW) and the Kardar-Parisi-Zhang (KPZ)
universality classes~\cite{EW,KPZ}. These classes describe the asymptotic scaling behavior
of the roughening interface and are characterized
by a certain set of exponents and scaling functions~\cite{fami85}.
In the corresponding Langevin equations, the KPZ class differs from the EW
class by a non-linear term that breaks reflection symmetry in height
direction. In experimental setups, where this symmetry is generally broken,
KPZ behavior is expected to be generic while linear growth (EW behavior) can be considered as
a special case.

Non-equilibrium wetting is usually modelled as a stochastic growth process 
on top of a hard-core substrate at height zero. Varying the growth rate the
presence of a substrate induces a wetting transition from a bound to a moving phase.
Concerning the scaling properties of the interface the substrate plays 
the role of a boundary: It does not change the universality class of the 
growth process itself, instead it imposes additional features. More specifically,
it gives rise to an additional order parameter and an associated critical exponent.
The situation is similar as e.g. in the Ising model, where a boundary 
induces an additional surface critical exponent. We will refer to these 
extended universality classes as the \textit{bounded} Edwards-Wilkinson (bEW) 
and the \textit{bounded} Kardar-Parisi-Zhang (bKPZ) class. For KPZ-type growth, however,
where the reflection symmetry is broken, it turns out that the new exponent also depends on the sign 
of the non-linear term. Therefore, one has to distinguish two different bounded KPZ 
classes, which we shall denote by the acronyms bKPZ+ and bKPZ-- according to the sign of
the nonlinear term.

\begin{figure}
\begin{center}
\includegraphics[width=85mm]{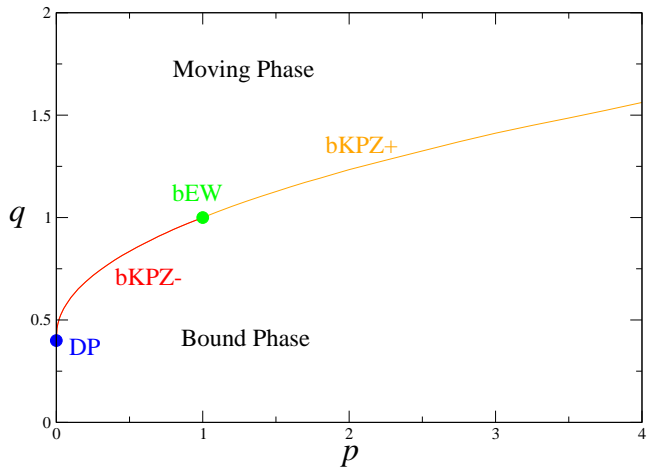}
\caption{Phase diagram of the wetting model with $r=1$}
\end{center}
\label{gphase1}
\end{figure}

One of these models for non-equilibrium wetting, which has been studied intensively
in the past, was introduced a decade ago in Ref.~\cite{hinr97}. 
It is a restricted solid-on-solid (RSOS) growth process on a $1$-dimensional  
lattice (see Sect. 2 for details), where the substrate is introduced by imposing the
condition that all heights have to be non-negative. The phase diagram of this model is shown in
Fig. \ref{gphase1}. Depending on the growth rate $q$ and the evaporation rate $p$,
the model exhibits a wetting transition from a bound to a moving phase. 
For $p=1$ the transition was shown to belong to the bEW universality
class, while for $p\neq1$ the transition belongs to one of the two bounded
KPZ classes. However, the reported estimates for the critical exponents are
still contradictory. One aim of the present paper is to clarify this issue 
and to confirm KPZ scaling along the whole line except for $p=0$ and $p=1$.

As shown in the figure, the transition line ends at the left terminal point
at $p=0$ and $q_c^{DP}= 0.3993(1)$. In this point the wetting model reduces to
a special growth process which was studied earlier by Alon \textit{et~al.}~\cite{alon96}
and has the special property that evaporation from completed layers is forbidden.
This means that the interface cannot have a negative growth velocity and therefore 
the presence of a wall makes no difference. 
Varying $q$ while keeping $p=0$ the model displays a roughening transition at $q=q_c^{DP}$.
It was shown that the dynamics of sites at the bottom layer can be related to a directed
percolation (DP)~\cite{hinr00} process, which is another class of non-equilibrium phase
transitions different from both bEW and bKPZ. Extending this analogy, it was
argued that the dynamics of the first few layers may be described in terms
of unidirectionally coupled growth processes~\cite{taub98,gold99}.

An open question, which will be addressed in the present work, concerns the
crossover from DP to the bKPZ-- class in the vicinity of the DP point. 
In order to describe this crossover, we propose to interpret evaporation from
the middle of a plateau with a small rate $p \ll 1$ as a weak external field in the
language of DP. This allows us to express the crossover exponent, which determines the
characteristic shape of the transition line as it approaches the DP point,
in terms of the response exponent of directed percolation.

The paper is organized as follows. In Section 2 we estimate the critical exponents for the model for a wetting transition by off-critical, time-dependent, and finite-size simulations. Moreover, we summarize a scaling picture for the bKPZ$\pm$ and bEW classes. Section 3 is devoted to the crossover from DP class ($p=0$) to bKPZ-- class ($0<p<1$) in the vicinity of the left terminal point of the transition line. Finally we demonstrate that the proposed interpretation of the wetting model as a DP process with a external field is in agreement with  numerical results.

\section{Estimation of the critical exponents}

\subsection{Definition of the model}   

The model for non-equilibrium wetting proposed in Ref.\cite{hinr97} is defined on a one-dimensional
lattice with $L$ sites and periodic boundary conditions. Each site $i$ is associated with a variable $h_i= 0,1,2,3...$ which describes the height of the interface at site $i$. The interface obeys the restricted solid-on-solid (RSOS) condition
\begin{equation}
|h_i-h_{i\pm 1}|\le 1,
\end{equation}
i.e., the heights at neighboring sites may differ by at most one unit. 

\begin{figure}
\begin{center}
\includegraphics[width=65mm]{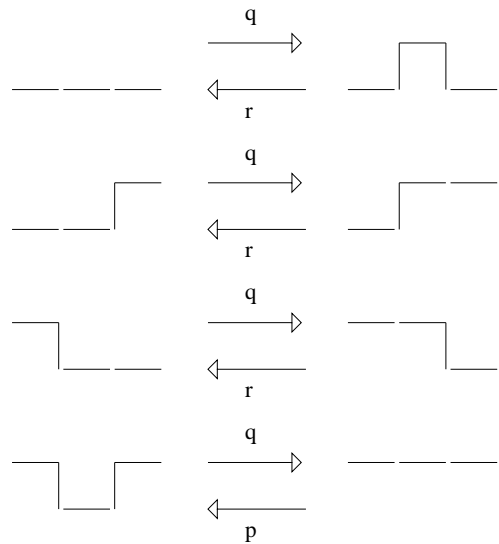}
\caption{Transition rates for the wetting model}
\label{frates1}
\end{center}
\end{figure}

The interface evolves in time by random-sequential updates as follows. For each update
a site $i$ of the lattice is randomly chosen and one of the following processes is selected (cf. Fig. ~\ref{frates1}): 
\begin{itemize}
\item[(a)] deposition of a particle ($h_i\rightarrow h_i+1$) with rate $q$,
\item[(b)] evaporation of a particle ($h_i\rightarrow h_i-1$) at the edges of plateaus with rate $r$,
\item[(c)] evaporation of a particle ($h_i\rightarrow h_i-1$) from the middle of a plateau with rate $p$.
\end{itemize}
A move is rejected if it would violate the RSOS condition. Moreover, the substrate is introduced by imposing the restriction that evaporation at zero height is forbidden. Each run starts with a flat interface at zero height. Without loss of generality we set $r=1$.

The phase diagram of the model is shown in Fig.~\ref{gphase1}. The transition is controlled by the growth rate $q$. Above the critical line the model is in the wet or moving phase, where the interface roughens and propagates at constant velocity. Below the critical line the interface remains bound and fluctuates close to the wall. The value of the evaporation rate $p$ determines the type of the phase transition. From the physical point of view small values of $p$ are more realistic since evaporation at the edges of a plateau is usually more likely than in the middle.

The order parameter for the wetting transition is the density of sites at zero height $n_0$. In the bound phase near the transition line $n_0$ goes to zero as
\begin{equation}
n_0\sim(q_c-q)^{\beta}\,,
\label{eqbeta}
\end{equation}
where $q_c(p)$ is the transition point and $\beta$ is a critical exponent. Similarly, the interface width $w$ diverges at the transition as
\begin{equation}
w\sim(q_c-q)^{-\zeta},
\label{eqzeta}
\end{equation}
where $\zeta$ is another critical exponent. 

The vertical line $p=0$ is special in so far as a layer, once completed, cannot evaporate again. As already mentioned in the introduction, this special case was studied in Ref.~\cite{alon96} and the exponent $\beta$ was found to be in agreement with the DP exponent $\beta= 0.27649(4)$~\cite{marr99}. Moreover, it was shown numerically that for $p=0$ the interface width diverges logarithmically near the critical point~\cite{hinr03}, which is consistent with the critical exponent $\zeta=0$. 

Another special case is $p=1$, where the wetting transition belongs to the bounded Edwards-Wilkinson (EW) class. In this case the stationary state in the bound phase can be computed exactly by transfer matrix methods \cite{hinr97}, which allows one to calculate the critical exponents $\beta= 1$ and $\zeta= 1/3$. 

For $0<p<1$ the model shows a different critical behavior as in the equilibrium case $p=1$. The first estimates reported in~\cite{hinr97}, using off-critical numerical simulations at $p=0.05$, are $\beta=1.51(6)$ and $\zeta=0.41(3)$. The purpose of this section is to revise these values, confirming the conjecture that in this regime the transition belongs to the bKPZ-- class. Similarly, for $p>1$ the transition is expected to belong to the bKPZ+ class.

\subsection{Off-critical simulations}   

\begin{figure}
\begin{center}	
\includegraphics[width=85mm]{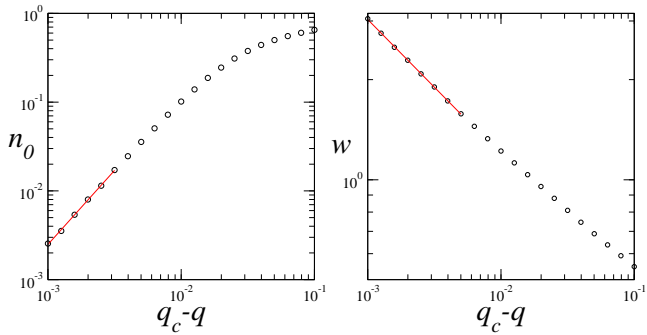}
\caption{Off-critical simulations. Density of sites with height zero $n_0$ (left) and the interface width $w$ (right)
as functions of the distance from the critical point $q_c= 0.4295(1)$. With $p=0.001$, $L=4096$ and $100$ independent realizations.} 
\label{goff}
\end{center}
\end{figure}

First we calculate the exponents $\beta$ and $\zeta$ by means of off-critical simulations. To reasons to be explained below, we use a very small value $p=0.001$. From the graphs shown in Fig.~\ref{goff} we obtain 
\begin{equation}
\beta= 1.67(5)\,,\qquad \zeta=0.41(5)\,.	
\end{equation}
We believe that this estimate of $\beta$ is larger than the one obtained previously in~\cite{hinr97} because of a crossover from EW to KPZ behavior. This crossover is known to be notoriously slow and may cause the impression as if the critical exponents depended continuously on $p$, varying from the EW exponent $\beta=1$ and some values larger than one. However, it seems that such an estimate is just an effective exponent measured in the crossover regime. As one approaches the critical line for $p<1$ and increases the numerical effort the effective exponent grows and slowly converges to the ¸`true' KPZ exponent. This crossover from EW to KPZ is expected to become more pronounced if we move away from the equilibrium case $p=1$. For instance, with simulations at $p=0.9$ we would obtain the effective exponent $\beta= 1.02(5)$. That is why we chose such a small value for $p$.

\subsection{Finite-size simulations}

\begin{figure}
\begin{center}
\includegraphics[width=85mm]{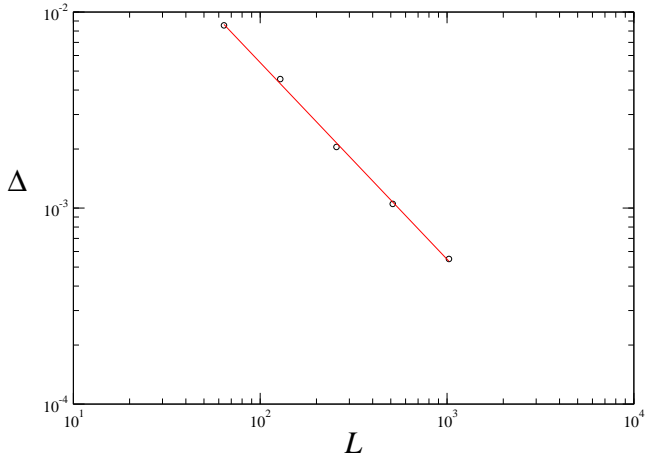}
\caption{Finite-size simulations. The figure shows the difference $\Delta$ between the finite-size critical point $q_c(L)$ and the extrapolated critical point $q_c(\infty)= 0.4295(3)$ as a function of $L$ for $p=0.001$.} 
\label{gfin}	
\end{center}
\end{figure}

According to the standard scaling picture of non-equilibrium phase transitions, the spatial correlation length $\xi_\perp$ near the critical point diverges as
\begin{equation}
\xi_\perp\sim (q_c-q)^{-\nu_\perp}.
\label{eqperp}
\end{equation} 
For $p=0$, where the model exhibits DP behavior, one obtains the DP values $\nu_\perp\approx 1.10$. For $0<p<1$, where the model is in the bKPZ-- class, we observe that the critical point, where the velocity of the free interface is zero, varies strongly with the system size. For instance, at $p=0.001$ for $L=128, 4096$ we found $q_c(L)= 0.425(1), 0.4295(1)$. Therefore, it is near at hand to postulate the relation
\begin{equation}  
q_c(\infty)-q_c(L)\sim L^{-1/\nu_\perp},
\end{equation}
where the $q_c(\infty)$ is the extrapolated value of the critical threshold. With $L= 64, 128, 256, 512, 1024$ we obtain 
$q_c(\infty)= 0.4295(3)$ and $\nu_\perp= 1.00(3)$. The data leading to this results, with 
$\Delta= q_c(\infty)-q_c(L)$ as a function of $L$, is shown in Fig. \ref{gfin}. We note that
the extrapolated value $q_c(\infty)= 0.4295(3)$ already coincides with the value $q_c(4096)= 0.4295(1)$ within error bars.

For finite growing interfaces, after an initial transient, the correlation length $\xi_\perp$ becomes of the same order as the system size. When this happens the interface width saturates. The value at which the interface width saturates depends on $L$ and scales as~\cite{bara95}
\begin{equation}
w_{sat}(L)\sim L^\alpha\,,
\label{eqalfa}
\end{equation}
where $w_{sat}$ denotes the saturation value of the interface width and $\alpha$ is the so-called roughness exponent. This relation is valid above the critical point and at the critical point for $p\neq0$. In one dimension one has
$\alpha= 1/2$ for both the KPZ and the EW class~\cite{bara95}. 
Eqs. (\ref{eqalfa}), (\ref{eqperp}), and (\ref{eqzeta}) imply the relation
\begin{equation}
\zeta= \nu_\perp\alpha.
\label{zeta}
\end{equation}
From $\nu_\perp= 1.00(3)$ we obtain $\zeta= 0.50(1)$, which is significantly larger than the numerical estimate in the previous subsection.

\subsection{Time-dependent simulations}

\begin{figure}
\begin{center}
\includegraphics[width=85mm]{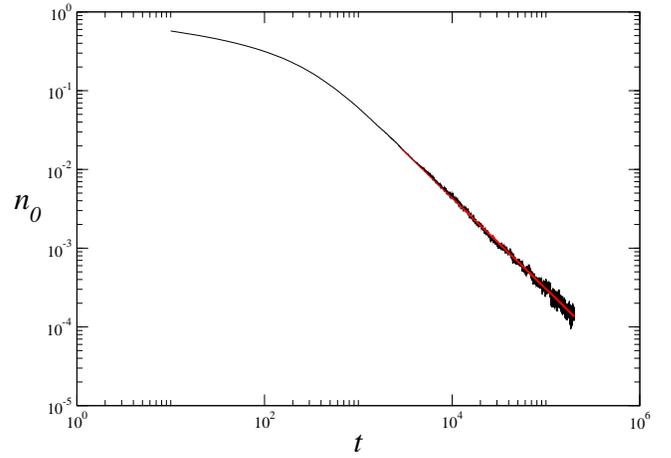}
\caption{Time-dependent simulations. The order parameter $n_0$ at the critical point $q_c= 0.4295(1)$ as a function of the number of Monte Carlo steps $t$ with $p=0.001$, $L=8192$ and $800$ independent realizations.} 
\label{gtim}
\end{center}
\end{figure}

Likewise, the temporal correlation length $\xi_\parallel$ diverges close to criticality as
\begin{equation}
\xi_\parallel\sim (q_c-q)^{-\nu_\parallel},
\label{eqparalelo}
\end{equation}
where $\nu_\parallel$ is the temporal critical exponent. From this relation and Eq.~(\ref{eqbeta})
we can conclude that the bottom layer density $n_0$ decays at criticality according to a power law
\begin{equation}
n_0\sim t^{-\theta}
\end{equation}
with 
\begin{equation}
\theta= \beta/\nu_\parallel.	
\end{equation}
The exponent $\theta$ was measured in Ref.\cite{kiss05}, simulating the so-called single step model which is known to belong to the bKPZ-- class. Our result $\theta=1.15(3)$, which is shown in Fig. \ref{gtim}, is compatible with their estimate $\theta= 1.184(10)$.  
%

The dynamical exponent $z= \nu_\parallel/\nu_\perp$, defined by 
\begin{equation}
\xi_\parallel\sim \xi_\perp^z,
\label{eqdin}
\end{equation}
for the KPZ class is $z=3/2$~\cite{bara95}. With $z=3/2$ and our previous estimate of $\nu_\perp$ we obtain $\nu_\parallel= 1.50(5)$. Together with $\theta= 1.15(3)$ this means that $\beta= 1.73(6)$.

We note that both values $\beta=1.73(6)$  and $\zeta= 0.50(2)$, obtained from finite-size and time-dependent simulations, are larger than the results obtained from off-critical simulations $\beta= 1.67(5)$ and $\zeta= 0.41(5)$. We believe that these discrepancies can be traced back to the fact that the EW-KPZ crossover is more influential in off-critical simulations.

\subsection{Scaling picture for equilibrium and non-equilibrium wetting}   

Let us now summarize the scaling picture and the values of the critical exponents. One-dimensional wetting models, as the one defined in Ref.~\cite{hinr97}, are characterized by four independent critical exponents, namely, three bulk exponents $\alpha,\nu_\perp,\nu_\parallel$ which take simple fractional values, and one exponent $\beta$ associated with the order parameter $n_0$, whose value is only known numerically. These exponents, together with related exponents $\zeta=\nu_\perp\alpha$ and $\theta=\beta/\nu_\parallel$, are listed in Table~\ref{t2}. The scaling hypothesis states that any quantities are invariant under the scaling transformation
\begin{eqnarray}
\vec{r}\to b\vec{r},\quad 
t\to b^z t,\quad
w \to b^\alpha w , \nonumber\\
n_0 \to b^{-\beta/\nu_\perp} n_0,\quad
\Delta \to b^{-1/\nu_\perp} \Delta\,,
\end{eqnarray}
where $b$ is a scaling factor and $\Delta=q-q_c$ denotes the distance from criticality.

\begin{table}[t]
\centering
\vspace{+0.1cm}
\begin{tabular}{|c|c|c|c|c|c|c|c|}
\hline
case 	       &$\alpha$&$z$  &$\nu_\perp$&$\nu_\parallel$&$\zeta$&$\theta$   & $\beta$ \\
\hline
DP	       & $0$  &$1.58$ &$1.10$& $1.73$& $0$  &$0.159$& $0.276$\\  
bKPZ--         &$1/2$   &$3/2$&$1$ 	  &$3/2$  	  &$1/2$  &$1.184(10)$& $1.776(15)$\\
bEW 	       &$1/2$   &$2$  &$2/3$      &$4/3$  	  &$1/3$  &$3/4$      & $1$\\
bKPZ+         &$1/2$   &$3/2$&$1$        &$3/2$  	  &$1/2$  &$0.228(5)$ & $0.342(8)$\\
\hline
\end{tabular} 
\caption{List of the critical exponents. Most of DP exponents in the frist line come from \cite{marr99}, the exceptions are $\alpha$ and $\zeta$ that come form \cite{hinr03}. The bEW and bKPZ exponents come from an exact result \cite{hinr97} and the best numerical estimatives for the exponent $\theta$ obtained in
\cite{kiss05} respectively combined with the scaling picture developed here.} 
\label{t2}
\end{table}

The three bulk exponents can be determined as follows.
For $p>0$ two of them, namely, the exponents $\alpha=\zeta/\nu_\perp$ and $z=\nu_\perp/\nu_\parallel$, are just the well-known bulk exponents of the EW or KPZ universality classes. To obtain the third exponent, let us consider the propagation velocity of a free interface. For $p\neq0$, the interface velocity varies linearly with the distance from the critical line, i.e.
\begin{equation}
v \sim q-q_c.
\label{eqvel}
\end{equation}  
Since the velocity is the temporal derivative of the mean height, which has the same scaling dimension as the width $w$, we expect that $w\sim (q-q_c)\xi_\parallel$. With equations (\ref{eqdin}) and (\ref{eqalfa}) we arrive at the scaling relation
\begin{equation}
\xi_\perp \sim (q-q_c)^{-1/(z-\alpha)},
\end{equation} 
hence
\begin{equation}
\nu_\perp= \frac{1}{z-\alpha}.
\label{eqsca}
\end{equation}  
This argument is hand-waving as it uses the properties of the free interface to predict the scaling properties of the wetting transition, but it is in agreement with all numerical observations. For example, for the KPZ case we have $\nu_\perp= 1$, in agreement with our numerical result, and for the EW case one obtains $\nu_\perp=2/3$, in agreement with an exact calculation~\cite{kiss05}. 

We note that the determination of $\beta$ (or likewise $\theta$) in the KPZ regime remains a numerically challenging task. Currently the most precise estimates come from the single-step model investigated in~\cite{kiss05}, reporting the values $\theta=1.184(10)$  for the bKPZ-- and $\theta= 0.228(5)$ for the bKPZ+ classes. The estimates obtained in the present simulations are consistent but not as precise, indicating that KPZ behavior of the model introduced in Ref.~\cite{hinr97} is not as `clean' as in the single-step model.

\section{The limit $p\to0$: Interpretation as a DP process in an external field}

As mentioned in the preceding section, the non-equilibrium wetting model introduced in Ref.~\cite{hinr97} includes a special case $p=0$, where it exhibits a transition belonging to the directed percolation (DP) universality class. The crossover from bKPZ-- to DP has not been studied so far.

The case $p=0$ is special in so far that a layer, once completed, cannot evaporate again. This means that the process does not feel the hard-core wall any more; it may be removed without changing the properties of the model. It was shown in Ref~\cite{alon96} that the sites at the actual bottom layer may be interpreted as the active sites of a DP process. This mapping is exact without RSOS condition but it remains effectively valid when the RSOS condition is imposed.

The key observation of the present work is that the process controlled by the parameter $p$, namely, evaporation from the middle of a plateau, corresponds to spontaneous creation of active sites in the language of DP. In DP such a spontaneous creation of activity is interpreted as an external field conjugate to the order parameter and the corresponding scaling laws are well understood. In this section we use this analogy to predict the properties of the crossover from DP to bKPZ-- in the non-equilibrium wetting process.

\subsection{Mapping the contact process in an external field to non-equilibrium wetting}

To understand the mapping between non-equilibrium wetting and DP in more detail, let us first consider the contact process (CP)~\cite{marr99} in an external field~$h$, which is defined by the following dynamical rules with random-sequential updates:
\begin{eqnarray}
1 \longrightarrow 0\qquad\textrm{ with rate}\qquad 1,\nonumber\\
0 \longrightarrow 1\qquad \textrm{with rate}\qquad h,\nonumber\\
01\,(10) \longrightarrow 11\,(11)\qquad\textrm{with rate}\qquad \lambda/2.
\label{rules1}
\end{eqnarray}
For $h=0$ one retrieves the usual CP  which exhibits a DP transition. For $h>0$ this transition is destroyed because the model does no longer have an absorbing state. 

Let us now compare this process with the unrestricted variant of the growth model introduced in~\cite{alon96}, which evolves according to the following dynamical rules:
\begin{eqnarray}
h_i \longrightarrow h_i+1 \qquad\textrm{with rate}\qquad 1,\nonumber\\
h_i \longrightarrow 0\qquad\textrm{with rate}\qquad h,\nonumber\\
h_i \longrightarrow min\{h_i,h_{i+1}\}\qquad\textrm{with rate}\qquad\lambda/2,\nonumber\\
h_i \longrightarrow min\{h_i,h_{i+1}\}\qquad\textrm{with rate}\qquad\lambda/2.
\end{eqnarray}
It is straightforward to verify that the variable $\eta_i=\delta_{h_i,0}$ follows exactly the dynamical rules given in (\ref{rules1}) and that the order parameter of the transition is the density of sites at the bottom layer. 

\begin{figure}
\begin{center}
\includegraphics[width=65mm]{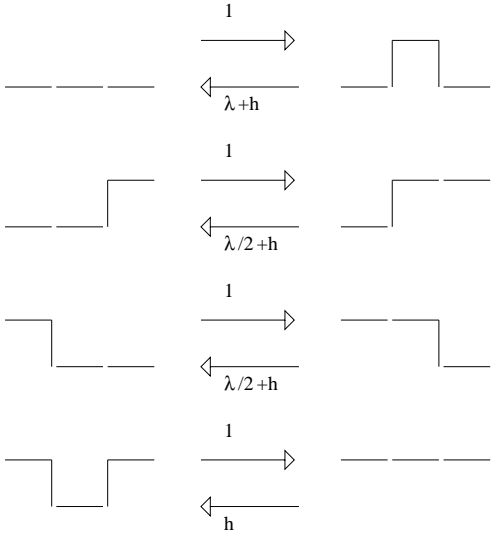} 
\caption{Transition rates for the restricted model}
\label{frates2}	
\end{center}
\end{figure}

For the restricted variant of the model introduced in~\cite{alon96} the above mapping is no longer exact. However, we argue that an external field $h$ can be introduced in the restricted model in an effective way by modifying the dynamical rules according to Fig. \ref{frates2}. This modification is motivated as follows. In the unrestricted version the likelihood for evaporation from the middle of a plateau does not depend on the actual configuration of the interface in the vicinity. To establish a similar independence in the restricted version, we introduce an additional evaporation rate $h$ which is the same for all interface configurations that respect the RSOS condition after an evaporation event. As usual, evaporation at zero height is forbidden.

Obviously, the dynamical rules listed in Fig. \ref{frates2} can be related to the wetting model by identifying the parameters
\begin{equation}
q=1,\qquad r=\lambda+h,\qquad p=h.
\end{equation}
Since $h\ge0$, this relation is valid only in the region $p\le r$. However, the mapping between the two models is not one-to-one because their transition rates are slightly different. While in the dynamical rules of Fig. \ref{frates2} evaporation with one neighbor at the same height happens with rate $\lambda/2+h$, the corresponding event in the wetting model takes place with rate $\lambda+h$. We expect that this minor difference does not change the critical behavior. Therefore, we conclude that the interpretation of evaporation in the middle of plateaus as an external field in the language of DP is still valid even in the restricted variant.

\subsection{Prediction of the form of the critical line in the limit $p\to 0$.}

We now demonstrate that the conjecture presented above can be used to predict the form of the phase transition line in the vicinity of the DP transition point $(p,q)=(0,0.3993(1))$. To this end we consider the stationary values of the order parameter $n_0$ along the line
\begin{equation} 
q= q_c^{DP}(1-p). 
\end{equation}
Using the parameter $h$ which is related to $q$ and $p$ by $q= 1/[(q_c^{DP})^{-1}+h]$ and $p= h/[(q_c^{DP})^{-1}+h]$ we define the field exponent $\delta$ by the asymptotic power law
\begin{equation}
n_0\sim h^{\delta},
\end{equation}
which is expected to be valid for small $h$. With numerical simulations we obtain the value $\delta= 0.121(5)$. This value differs from the DP exponent $\beta= 0.27649(4)$ measured in vertical direction along the line $p=0$. Therefore, approaching the DP point from different directions we find two different exponents. The smaller one (with a slower decay of $n_0$) is expected to dominate all other directions except the vertical one. This allows us to conclude that in horizontal direction, i.e., for $q= q_c^{DP}$ and small values of $p$, the order parameter vanishes as
\begin{equation}
n_0\sim p^{\delta}.
\end{equation}
This equation together with equation (\ref{eqbeta}) implies
\begin{equation}
q_c-q_c^{DP}\sim p^{y},
\end{equation}
where $y= \delta/\beta= 0.44(2)$ is a crossover exponent. 

\begin{figure}
\begin{center}
\includegraphics[width=85mm]{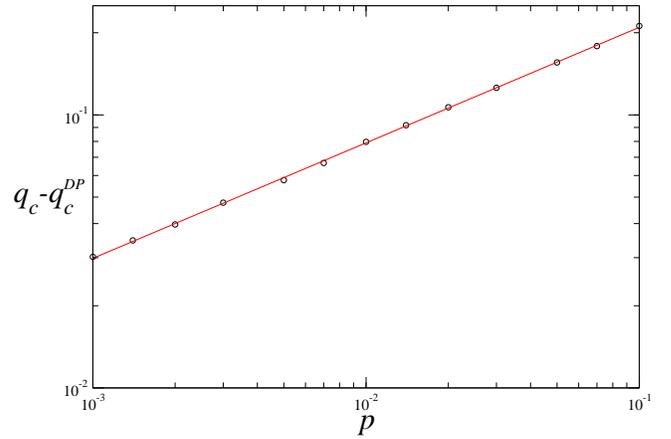}
\caption{The difference $q_c-q_c^{DP}$ as a function of $p$.} 
\label{goff2}
\end{center} 
\end{figure}

The crossover exponent describes how the critical line approaches the DP point in the sense that it determines the critcal line concavity near the DP point. In fact, plotting $q_c-q_c^{DP}$ versus $p$ in a double-logarithmic plot (see Fig.~\ref{goff2}) one obtains a straight line with the slope $0.43(2)$, which coincides with the value of~$y$. This means that the interpretation of evaporation from the middle of a plateau as an external field in the language of DP yields the correct crossover exponent describing the curvature of the phase transition line.

\subsection{Critical behavior of the first few layers}

So far we considered only the density of sites at the bottom layer $n_0$. Similarly one can study the density of sites $m_k$ whose heights are less or equal than $k$. As shown in~\cite{taub98,gold99}, these order parameters vary in vertical direction ($p=0$) by power laws $m_k \sim (q_c^{DP}-q)^{\beta_k}$ with individual exponents $\beta_0,\beta_1,\beta_2,\ldots$, where $\beta_0=\beta$ is the ordinary density exponent of DP. In the same way we can now define the exponents $\delta_k$ by 
\begin{equation}
m_k\sim h^{\delta_k}\sim p^{\delta_k}\,,
\end{equation} 
where $\delta_0$ is just the exponent $\delta$ of the preceding subsection.

\begin{figure}
\begin{center} 
\includegraphics[width=85mm]{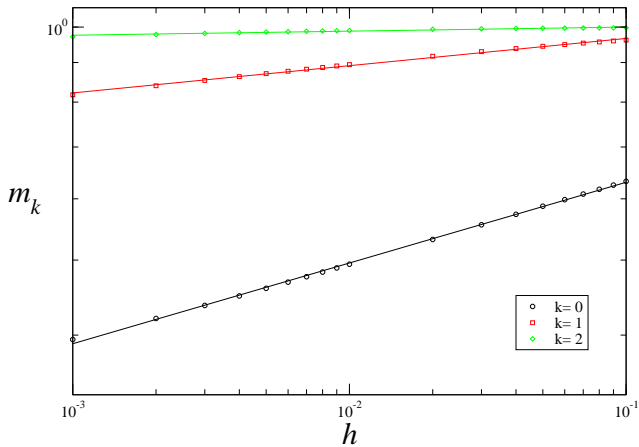}
\caption{$m_k$ as a function of $h$ with $\lambda_c= 4.30195(5)$~\cite{hinr03} for the restricted model with the external field. In this case $L= 4096$ and $1000$ is the number of independent realizations.}
\label{ghr}
\end{center}
\end{figure}

As an example, Fig.~\ref{ghr} shows numerical measurements of $m_k$ for the restricted model. The estimates of $\delta_k$ are listed in Table \ref{tabledelta}. All these values are in fair agreement with the numerical estimates for the unidirectionally coupled DP reported in~\cite{gold99}. 

\begin{table}[b]
\centering
\vspace{+0.1cm}
\begin{tabular}{|c|c|c|c|}
\hline
model & $\delta_{0}$ & $\delta_{1}$ & $\delta_{2}$  \\
\hline
unrestricted & $0.107(2)$ & $0.040(3)$ & $0.014(2)$\\
\hline
restricted & $0.104(5)$ & $0.039(5)$ & $0.008(5)$\\
\hline
wetting & $0.121(5)$ & $0.047(5)$ & $0.009(5)$\\
\hline
\end{tabular} 
\caption{Numerical estimates of the exponents $\delta_{k}$}
\label{tabledelta}
\end{table}

\section{Conclusion}

In this paper we have presented improved estimates of the critical exponents for the non-equilibrium wetting model of Ref.~\cite{hinr97} in the parameter range $0<p<1$. We have derived relations between the exponents, in agreement with the numerical results, and used them to give a complete set of the exponents along the whole transition line (see Table \ref{t2}).

This work is focused on the case $0< p < 1$, where the wetting transition belongs to the bKPZ-- class. For $p>1$, where the transition belongs to the bKPZ+ class, we could do a similar analysis but we would have to use extremely large values of $p$ to overcome crossover effects, but numerical simulations turn out to be inefficient in this limit.

Presently the most reliable results for the critical exponent $\theta$ of the bounded KPZ classes were measured using the so-called single step model~\cite{kiss05}. This model has a moving wall and is always exactly at the critical point. However, the single step model does not allow one to perform off-critical simulations. This is the reason why the wetting model is more suitable to confirm the scaling relations discussed in this paper.  

The non-equilibrium wetting model introduced in~\cite{hinr97} has another interesting feature, namely, a special transition point at $p=0$, where the critical behavior belongs to the directed percolation universality class. Surprisingly, even for very small $p$, the critical behavior changes entirely. For example, the exponent $\beta$, which describes the density of sites at zero height, jumps from the DP value $\beta\approx 0.28$ for $p=0$ to a large value $\beta\approx 1.78$ of the bKPZ-- class. To our knowledge, this is the only case where the order parameter exponent $\beta$ is larger than one. This is due to the fact that the interface at criticality has only very few contact points where it touches the wall. This number is so small that the correlation length $\xi_\perp$ of the interface fluctuations becomes smaller than the average distance between two contact points~\cite{kiss05}, which is equal to the inverse of the density of sites at the bottom layer.   

The main result of this work is the conjecture that a wetting process for small $p$ can be interpreted  as a DP process in an external field. This conjecture allows us to interpret the crossover from the DP class to the bKPZ-- class as a small external field $h$ that eliminates the DP transition. Moreover, it is consistent with the scaling picture and can be confirmed by numerical simulations. Calculating the corresponding crossover exponent $y=\delta/\beta$, we can predict the shape of the critical line near $p=0$.    

\noindent \textbf{Acknowledgment:}\\
We thank the Deutsche Forschungsgemeinschaft for financial support (HI 744/3-1).

\end{document}